\pdfoutput=1
\documentclass[conference]{IEEEtran}
\usepackage[latin1]{inputenc}
\usepackage{times,amsmath}
\usepackage{amssymb}
\usepackage{pstool}
\usepackage{subfigure}
\usepackage{multirow}
\usepackage{enumerate}
\usepackage{graphicx}
\usepackage{MnSymbol}
\usepackage{stfloats} 
\usepackage[table]{xcolor}
\usepackage[square, comma, sort&compress, numbers]{natbib}
\usepackage{nohyperref}
\usepackage{epstopdf}
\usepackage{comment}
\usepackage[ruled,noend, vlined,linesnumbered]{algorithm2e}
\usepackage{amsthm}
\usepackage[T1]{fontenc}
\usepackage{soul}

\usepackage[T1]{fontenc} %%%%% Optional T1 Font Encoding %%%%%
\begin{document}

%\title{Non Line-of-Sight Prediction via Physical Layer Features: A Machine Learning based Approach}
%\title{Pathloss-based non Line-of-Sight Prediction in an Indoor Environment: An Experimental Study }
\title{Pathloss-based non-Line-of-Sight Identification in an Indoor Environment: An Experimental Study }

\author{\IEEEauthorblockN{
M. Asim\IEEEauthorrefmark{1}, 
M. Ozair Iqbal\IEEEauthorrefmark{1}, 
Waqas Aman\IEEEauthorrefmark{2}, 
M. Mahboob Ur Rahman\IEEEauthorrefmark{1},
Qammer H. Abbasi\IEEEauthorrefmark{3}}

\IEEEauthorblockA{
    \IEEEauthorrefmark{1} Electrical Engineering Department, Information Technology University, Lahore 54000, Pakistan \\
    \IEEEauthorrefmark{2}Division of Information and Computing Technology, College of Science and Engineering, \\Hamad Bin Khalifa University, Qatar Foundation, Doha, Qatar.\\
   \IEEEauthorrefmark{3} Department of Electronics and Nano Engineering, University of Glasgow, Glasgow, G12 8QQ, UK\\
waman@hbku.edu.qa, mahboob.rahman@itu.edu.pk,  Qammer.Abbasi@glasgow.ac.uk}}
%\author{M. Asim, M. Ozair Iqbal, Waqas Aman, M. Mahboob Ur Rahman and Qammer H. Abbasi}
\maketitle
\begin{abstract}
{

This paper reports the findings of an experimental study on the problem of line-of-sight (LOS)/non-line-of-sight (NLOS) classification in an indoor environment. Specifically, we deploy a pair of NI 2901 USRP software-defined radios (SDR) in a large hall. The transmit SDR emits an unmodulated tone of frequency 10 KHz, on a center frequency of 2.4 GHz, using three different signal-to-noise ratios (SNR). The receive SDR constructs a dataset of pathloss measurements from the received signal as it moves across 15 equi-spaced positions on a 1D grid (for both LOS and NLOS scenarios). We utilize our custom dataset to estimate the pathloss parameters (i.e., pathloss exponent) using the least-squares method, and later, utilize the parameterized pathloss model to construct a binary hypothesis test for NLOS identification. Further, noting that the pathloss measurements slightly deviate from Gaussian distribution, we feed our custom dataset to four machine learning (ML) algorithms, i.e., linear support vector machine (SVM) and radial basis function SVM (RBF-SVM), linear discriminant analysis (LDA), quadratic discriminant analysis (QDA), and logistic regression (LR). It turns out that the performance of the ML algorithms is only slightly superior to the Neyman-Pearson-based binary hypothesis test (BHT). That is, the RBF-SVM classifier (the best performing ML classifier) and the BHT achieve a maximum accuracy of 88.24\% and 87.46\% for low SNR, 83.91\% and 81.21\% for medium SNR, and 87.38\% and 86.65\% for high SNR.
%\textcolor{blue}{We also investigate the case of multiple features (i.e., path loss, AoA, ToA, RSS, etc.) and show that multiple-feature-based ML algorithms perform much better than single-feature-based ML algorithms and the hypothesis tests}. 

}
\end{abstract}

\begin{IEEEkeywords}
line-of-sight (LOS), non-line-of-sight (NLOS), classification, least-squares, binary hypothesis test, machine learning, support vector machine.
\end{IEEEkeywords}

\section{Introduction}

The upcoming 6G cellular standard aims to provide an immersive and personalized user experience by enabling a wide range of novel location-based applications, including augmented reality (AR), virtual reality (VR), and mixed reality (MR). To this end, precise indoor localization is the pre-requisite to realize such applications, which will allow seamless integration of virtual and physical environments, enable precise positioning of virtual objects, and deliver context-aware services to the users \cite{shen2023five}. 

Indoor localization is a challenging task due to the lack of global positioning system (GPS) signals indoors, due to the presence of obstacles/blockages, multipath, and random signal variations due to rich scattering in indoor environments.
To date, numerous indoor propagation models and various methods for indoor localization have been reported in literature to examine and to undo the impact of non-idealities (e.g., multi-path, blockages) \cite{zafari2019survey}. Some popular methods for indoor localization include the following: fingerprinting (scene analysis) based, time of arrival (ToA) based, angle of arrival (AoA) based, phase of arrival (PoA) based, time of flight (ToF) based, time difference of arrival (TDoA) based, and received signal strength (RSS) based, Ricean k-factor based \cite{zafari2019survey}. 

This work focuses on the challenge posed by the blockages to the indoor localization systems. Specifically, blockages turn a link into a non-line-of-sight (NLOS) link, which in turn makes the distance/AoA estimates obtained by the indoor localization algorithms biased. Thus, NLOS conditions when exist, degrade the accuracy of the indoor positioning systems due to the ranging errors. Therefore, accurate NLOS prediction/classification is the need of the hour. NLOS prediction helps indoor positioning systems identify and mitigate the effects of NLOS conditions, and thus could lead to a boost in the accuracy of the indoor position estimates \cite{nessa2020survey}. Other than indoor localization, NLOS identification could also help solve many other important problems, e.g., it could help discover blocked THz links indoors, which might prompt a THz access point to provide service to the associated users by means of a reconfigurable intelligent surface (RIS) panel, therefore, improving the coverage of the indoor THz link \cite{aman2023downlink}. NLOS identification, thus, provides valuable insights for the design of blockage-aware user association algorithms and handover management algorithms.

The problem of NLOS identification has recently caught attention by the research community, and a number of works have been reported in the literature, to date. Thus, the discussion of the selected related works is in order. \cite{li2017nlos} utilizes a WiFi system to collect channel frequency response (CFR) and channel impulse response (CIR) samples, extracts a number of statistical features (e.g., mean, variance, skew, kurtosis, etc.) from the fine-grained channel state information (CSI) and feeds them to a support vector machine that does the NLOS identification.
Authors of \cite{fan2019non} consider an ultra-wideband system and use a semi-supervised learning approach, i.e., they utilize the expectation maximization algorithm to learn the parameters of their Gaussian mixture model for NLOS identification. %The k-factor (if greater than 0dB) of the rician distribution is an indicator of the LOS condition. 
The work \cite{Yu:TVT:2009} extracts a number of features (e.g., AoA, ToA, RSS, etc.) from the incoming received signal and utilizes various methods (e.g., Neyman-Pearson method) from the statistical decision theory in order to identify the NLOS conditions.
The authors in  \cite{Xiao:TWC:2015} collect RSS samples using an indoor WiFi system and extract multiple statistical features from the RSS time series in order to feed them to a least squares support vector machine and to a hypothesis test which do NLOS identification. They further do NLOS mitigation by designing various distance estimation algorithms under both line-of-sight (LOS) and NLOS conditions.

Recently, a few researchers have proposed machine learning (ML) and deep learning methods for NLOS identification. 
For example, the authors in \cite{Choi:TVT:2018} implement a recurrent neural network (RNN) model that utilizes the CSI measurements collected in an indoor office environment, in order to identify the NLOS condition. \cite{sang2020identification} studies the problem of ultra-wideband based wireless ranging, and utilizes a support vector machine, a random forest classifier and a multi-layer perceptron to solve the three-class classification problem with the following classes: LOS, NLOS, and multipath. The authors in \cite{che2022feature} propose feature-based Gaussian distribution method and generalized Gaussian distribution method for NLOS detection under the constraint of an imbalanced dataset (with very few examples from the NLOS class). The authors in \cite{si2023lightweight} propose a novel algorithm for LOS/NLOS classification based on a multi-layer perceptron that utilizes both manually extracted features as well as the features obtained from a convolutional neural network (CNN) using raw CIR inputs. Last but not the least, the authors in \cite{SINGH2023102118} study the problem of localization in a millimeter-wave wireless communication system, and train and test a two-stage unsupervised ML model on CSI data in order to classify LOS/NLOS.

On the prototyping front, there are a handful of works that report experimental results on indoor localization \cite{robesaat2017improved,Gonzalez:Access:2019,Marter:IOT:2023}. For example, the authors in \cite{robesaat2017improved} use a bluetooth low energy (BLE) module to do indoor localization via different approaches, i.e., trilateration, dead reckoning, and the fusion method. Further, an experimental study that investigates the relation between the accuracy and energy consumption in a WiFi fingerprinting-based indoor localization system is proposed in \cite{Gonzalez:Access:2019}. Finally, ML-assisted indoor localization is discussed in \cite{Marter:IOT:2023} where support vector regression (SVR) is done on CFR measurements obtained via a BLE module, in order to accomplish indoor localization in a multipath environment.  
%\textcolor{blue}{Nevertheless, to the best of the authors' knowledge, this problem has not been studied before. }

{\bf Contributions.}
This is an experimental study where we do an extensive data collection campaign via a pair of NI 2901 USRP software-defined radios in order to collect pathloss measurements in 5G FR1 band in an indoor setting. We first apply a least-squares method on the pathloss measurements in order to parameterize the pathloss model which is later utilized to construct a Neyman-Pearson-based binary hypothesis test.
Further, noting that the pathloss measurements slightly deviate from the Gaussian distribution, we apply following four machine learning algorithms to the experimental data collected: linear support vector machine (SVM) and radial basis function SVM (RBF-SVM), linear discriminant analysis (LDA), quadratic discriminant analysis (QDA), and logistic regression (LR). It turns out that the performance of the best-performing ML algorithm (i.e., RBF-SVM) is only slightly superior than its counterpart from statistical decision theory, i.e., binary hypothesis test. 
 
%\textcolor{blue}{We also investigate the case of multiple features (i.e., pathloss, AoA, ToA, RSS, etc.) and show that multiple-feature-based ML algorithms perform much better than single-feature-based ML algorithms and the hypothesis tests. }

{\bf Outline.} The rest of this paper is organized as follows. Section II describes the experimental setup and the data collection process. Section III presents the two proposed methods for NLOS identification in detail. Section IV provides some selected results. Section V concludes the paper. 
 
%\textcolor{red}{self-notes: testing on different test data? NLOS mitigation as well via adjusting the distance estimate? log-distance model for pathloss? and a hypothesis test based on that? test data should be collected at different points/receiver locations? what is the difference between pathloss and RSS? what about the kurtosis of pathloss histogram? does derived features (from pathloss) contain further new information/innovation to help us in decision making? why did we choose these particular ML algorithms (SVM, LDA, QDA, LR)?}

\section{Experimental Setup \& Data Collection}

\begin{figure}[h]
\centering{\includegraphics[width=90mm]{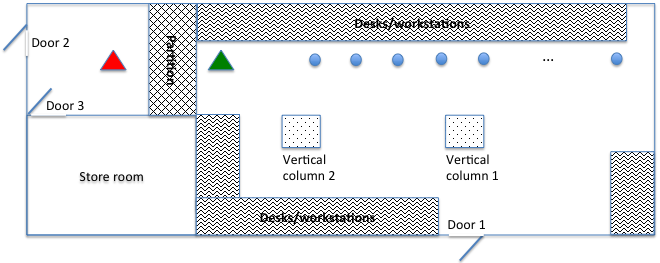}}
\caption{The experimental setup (not to scale). The receiver (blue circle) is placed at 15 different positions on a 1D grid. The transmitter is either in LOS of the receiver (green triangle), or, in NLOS condition (red triangle).}
\label{fig:exp-setup}
\end{figure}

We performed our data collection experiments in 5G FR1 band by deploying a pair of NI 2901 USRP software-defined radios (SDR) in one of the research labs at the Information Technology University (ITU), Lahore, Pakistan. The detailed layout of the room where we conducted our experiments is shown in Fig. \ref{fig:exp-setup}.  

As can be seen in Fig. \ref{fig:exp-setup}, the receiver was placed at $P=15$ different positions on a linear grid with inter-position spacing of 60 cm. The minimum transmit-receive spacing is 125 cm as per (10 wavelengths) requirement for the receiver to be in far field of the transmitter, while the maximum transmit-receive spacing is 900 cm. Directional (Horn) antennas with a maximum gain of 20 dB each were used at both ends (this helped reduce the impact of multipath for the LOS measurements). Center frequency $f_c$ was set to 2.4 GHz (i.e., the ISM band), while the sampling rate of both the transmit and the receive SDR was set to 200K samples/s. For both LOS and NLOS scenarios, measurements were taken for three different signal-to-noise ratio (SNR) conditions by changing the normalized amplitudes $A_t$ of the transmitted signal in the following range: 0.4,0.5,0.6. The transmit node sent a unmodulated tone of frequency 10 KHz. The channel was considered to be time-slotted with a slot length of 10 ms (large enough so that all the multipath components could be lumped together in one slot). The received signal directly provided the instant RSS measurements. Subsequently, the instant RSS sample within a timeslot were averaged to get a more stable and reliable RSS estimate. Averaging also helped us get rid of the small-scale fading occurring on a relatively fast time-scale. The averaged RSS measurements were then translated into the pathloss measurements using the Friis equation assuming that the antenna gains on both ends as well as the transmit power is known. That is, pathloss $=\frac{P_t}{P_r}$ where $P_t$ is the known transmit signal power, while $P_r=(\text{RSS})^2$ is the received signal power. The pathloss measurements were then used to construct a least-squares (LS) problem where the pathloss exponent $\alpha$ was computed for both LOS and NLOS scenarios, for each of three link conditions. A total of $N=5000$ measurements were obtained for each of the 15 receiver positions for both LOS and NLOS scenarios (in order to construct a balanced dataset), for three different SNR values. 

\begin{figure}[h]
\centering{\includegraphics[width=95mm]{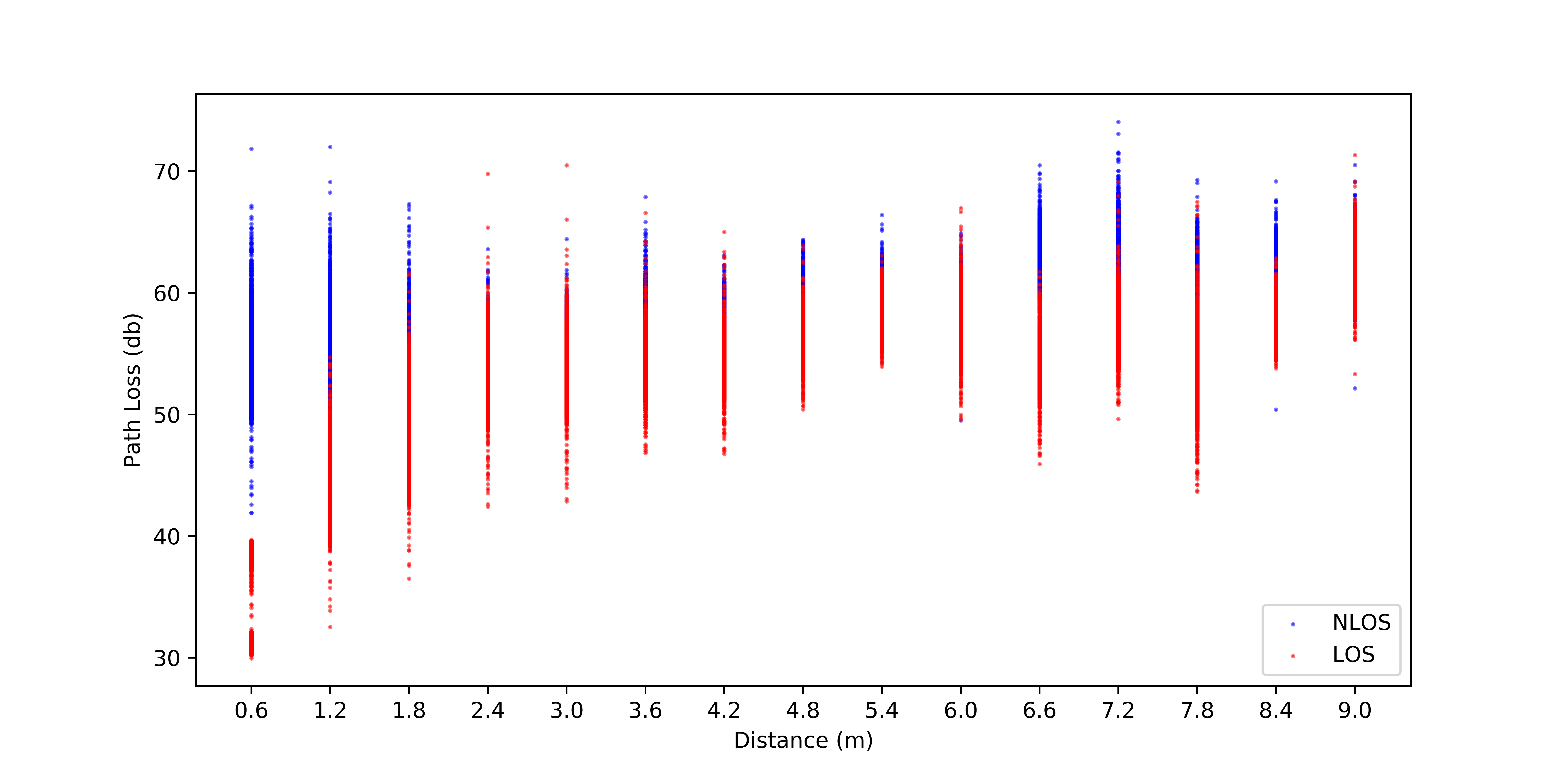}}
\caption{Pathloss measurements obtained via our experimental setup consisting of two NI 2901 USRP SDRs when the receive SDR moves across 15 equi-spaced positions on a 1D grid.}
\label{fig:pathlossmeas}
\end{figure}

{\it Feasibility of pathloss as core feature for NLOS identification.} Fig. \ref{fig:pathlossmeas} plots the pathloss measurements that we obtained by moving the SDR receiver on a 1D grid during our data collection campaign, for both LOS and NLOS scenarios. Fig. \ref{fig:pathlossmeas} attests to the fact that the pathloss (exponent $\alpha$) is higher for the NLOS scenario, compared to the LOS scenario (as is well-known in the literature). 

\section{The Proposed Methods}

We first describe our binary hypothesis test for NLOS identification in detail. We then discuss the essentials of the four machine learning classifiers that we have implemented for NLOS identification. 

\subsection{NLOS Identification via Binary Hypothesis Testing}

The binary hypothesis testing method for NLOS identification requires the measurements of pathloss conditioned on the two hypotheses, i.e., LOS and NLOS. Therefore, we first present a least-squares method for the estimation of the pathloss parameters. We then design a binary hypothesis test for NLOS identification and compute the two error probabilities (i.e., false alarm rate and missed-detection rate).

\subsubsection{Least-Squares Estimation of the Pathloss Parameters}
The Friis equation is: $P_r = P_t G_t G_r \bigg(\frac{\lambda}{4\pi d} \bigg)^\alpha$ where $P_r$ is the received power, $P_t$ is the transmit power, $G_t$ is the transmit antenna gain, $G_r$ is the receive antenna gain, $\lambda=\frac{c}{f_c}$ is the wavelength, $c$ is the speed of light, $f_c$ is the center frequency, $d$ is the separation between the transmitting node and the receiving node, $\alpha$ is the pathloss exponent. Re-arranging Friis equation, we obtain the following distance-dependent pathloss model:
\begin{equation}
PL(d) = \frac{P_t}{P_r} = \frac{1}{G_t G_r} \bigg(\frac{4\pi d}{\lambda} \bigg)^\alpha
\end{equation}
Equivalently, in dB scale, we have:
\begin{equation}
PL_{dB}(d) = \mathcal{A} + \alpha10\log_{10}B(d) 
\end{equation}
where $\mathcal{A} = -10\log_{10}(G_tG_r)$ and $B(d)=\frac{4\pi d}{\lambda}$.

As mentioned earlier, in this work, we collect noisy measurements of instant RSS, square them to translate them into instant $P_r$ measurements which are further translated into pathloss measurements by multiplying $1/P_r$ with the (known) $P_t$. Then, the least-squares (LS) estimate of $\mathcal{A}$ and $\alpha$ is: $\mathbf{\Theta}=\mathbf{X}^T(\mathbf{XX}^T)^{-1}\mathbf{y}$ where $\mathbf{y}\in \mathrm{R}_+^{(N\times P)\times 1}$ represents the measurement vector containing pathloss values, $\mathbf{\Theta}=[\mathcal{A},\alpha]^T \in \mathrm{R}_+^{2\times 1}$ is the vector of unknowns, $\mathbf{X}=[\mathbf{x},\mathbf{1}_{(N\times P)\times 1}]\in \mathrm{R}_+^{(N\times P)\times 2}$ is the system matrix, $\mathbf{x}=[\mathbf{x}^{(1)}_N,...,\mathbf{x}^{(P)}_N]^T$, $P$ is the number of receiver positions, $N$ is the number of measurements obtained at each receiver position. 

Table \ref{table:alpha} summarizes the vector of unknowns $\mathbf{\Theta}$ estimated via the LS method for both LOS and NLOS scenarios, for three different SNRs.

\begin{center}
\begin{table}[h]
\caption{Pathloss parameters estimated via least-squares method}
    \begin{tabular}{ | l | l | l | l | p{1.5cm} |}
    \hline
    link condition & $\alpha_{LOS}$ & $\mathcal{A}_{LOS}$ & $\alpha_{NLOS}$ & $\mathcal{A}_{NLOS}$ \\ \hline
    low SNR ($A_t=0.4$) & 1.04 & 50.02 & 1.24 & 53.51 \\ \hline
    medium SNR ($A_t=0.5$) & 1.68 & 44.89 & 1.09 & 52.20\\ \hline
    high SNR ($A_t=0.6$) & 1.75 & 42.53 & 0.87 & 52.93 \\
    \hline
    \end{tabular}
    
    \label{table:alpha}
\end{table}
\end{center}

\subsubsection{Binary Hypothesis Test}
With pathloss parameters in hand, we have the following binary hypothesis test (BHT) for NLOS identification (assuming Gaussian measurement noise):
\begin{equation}
	\label{eq:H0H1}
	 \begin{cases} 
	 H_0 (\text{LOS}): & {z} = \mathcal{A}_{los} + \alpha_{los}10\log_{10}B + n  \\ 
     H_1 (\text{NLOS}): & {z} = \mathcal{A}_{nlos} + \alpha_{nlos}10\log_{10}B + n 
     \end{cases}
\end{equation}
where $z$ is the pathloss measurement, $n\sim N(0,\sigma^2)$ is measurement error. Let $m_0=\mathcal{A}_{LOS} + \alpha_{LOS}10\log_{10}B$ and $m_1=\mathcal{A}_{NLOS} + \alpha_{NLOS}10\log_{10}B$. Then, $z|H_0\sim N(m_0,\sigma^2)$ and $z|H_1\sim N(m_1,\sigma^2)$. This translates to the following log-likelihood ratio test (LLRT):
\begin{equation}
	\label{eq:H0H1_1_2}
	 z \underset{H_0}{\overset{H_1}{\gtrless}} \delta = \bigg( \frac{\sigma^2\ln \eta}{m_1-m_0} + \frac{m_0+m_1}{2} \bigg) 
\end{equation}
where $\eta=\ln(\pi(0)/\pi(1))$.

Then, the probability of false alarm is given as: PFA $=Pr(z>\delta|H_0)=Q(\frac{\delta-m_0}{\sigma})$ where $Q(x)=\frac{1}{\sqrt{2\pi}} \int_x^\infty  e^{-\frac{t^2}{2}} dt$ is the standard $Q$-function. Next, the probability of detection is given as: PD$=1-$ PMD $=1-Pr(z<\delta|H_1)=Q(\frac{\delta-m_1}{\sigma})$.

\subsection{NLOS Identification via Machine Learning Classifiers}

\begin{figure}[h]
\centering{\includegraphics[width=90mm]{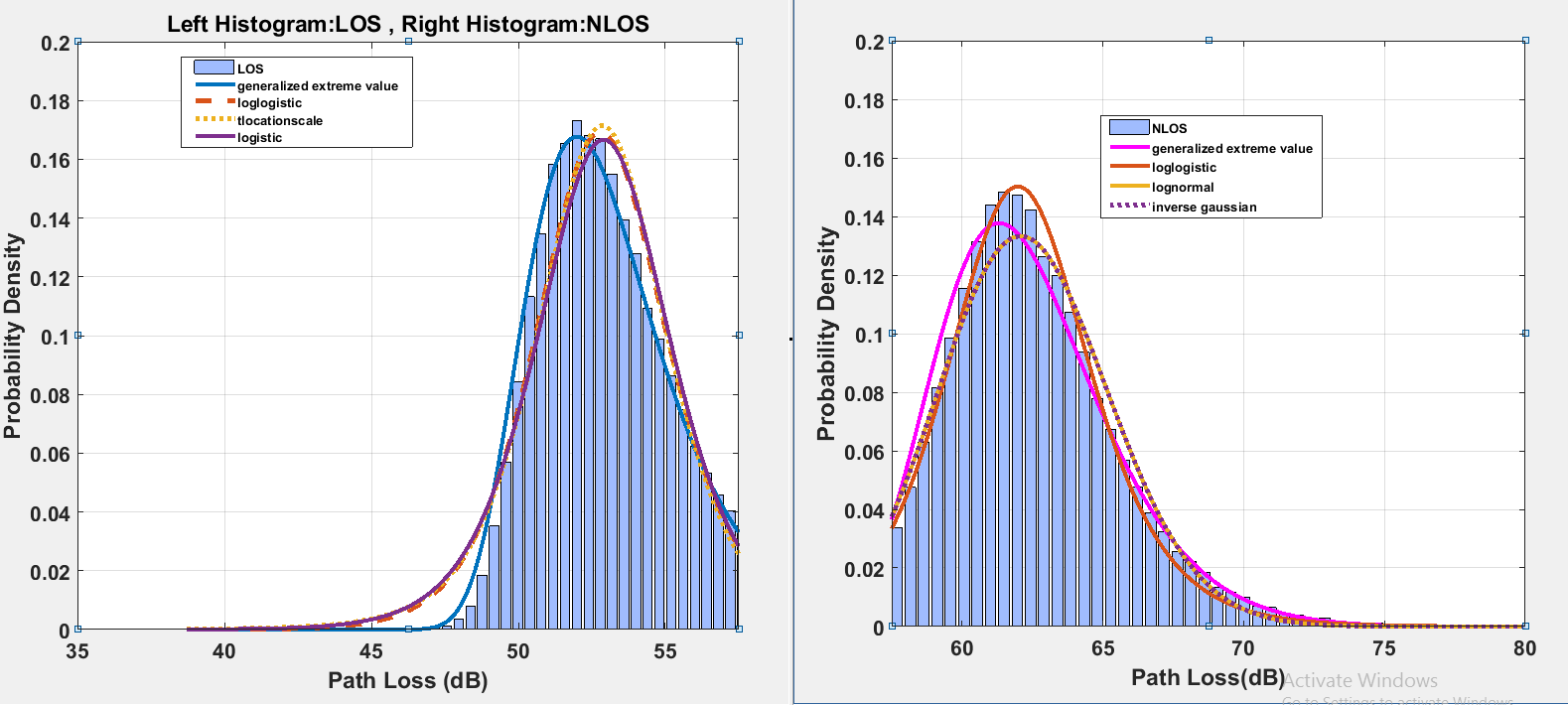}}
\caption{Pathloss histogram does not fits well to normal distribution (for both LOS and NLOS scenarios).}
\label{fig:pathlosshist}
\end{figure}

We note that the pathloss measurements collected in real-time setup via an SDR-pair slightly deviate from the Gaussian distribution (see Fig. \ref{fig:pathlosshist}). Therefore, the binary hypothesis test defined above that assumes Gaussian distribution for the measurement error may not work very well. However, it is well-known that the machine learning algorithms can cope with this situation (model mismatch) by learning the distribution from the training data.
Therefore, we implement the following machine learning algorithms in Python: linear support vector machine (SVM) and radial basis function SVM, linear discriminant analysis (LDA), quadratic discriminant analysis (QDA), and logistic regression (LR). We train and test the four ML classifiers on our custom dataset with a train-validation-test split of 70-15-15 (\%). 
%10-fold cross-validation is done.

\section{Results}

\begin{figure}[t]
\centering
\subfigure[low SNR ($A_t=0.4$) ]{
\includegraphics[width=.5\textwidth]{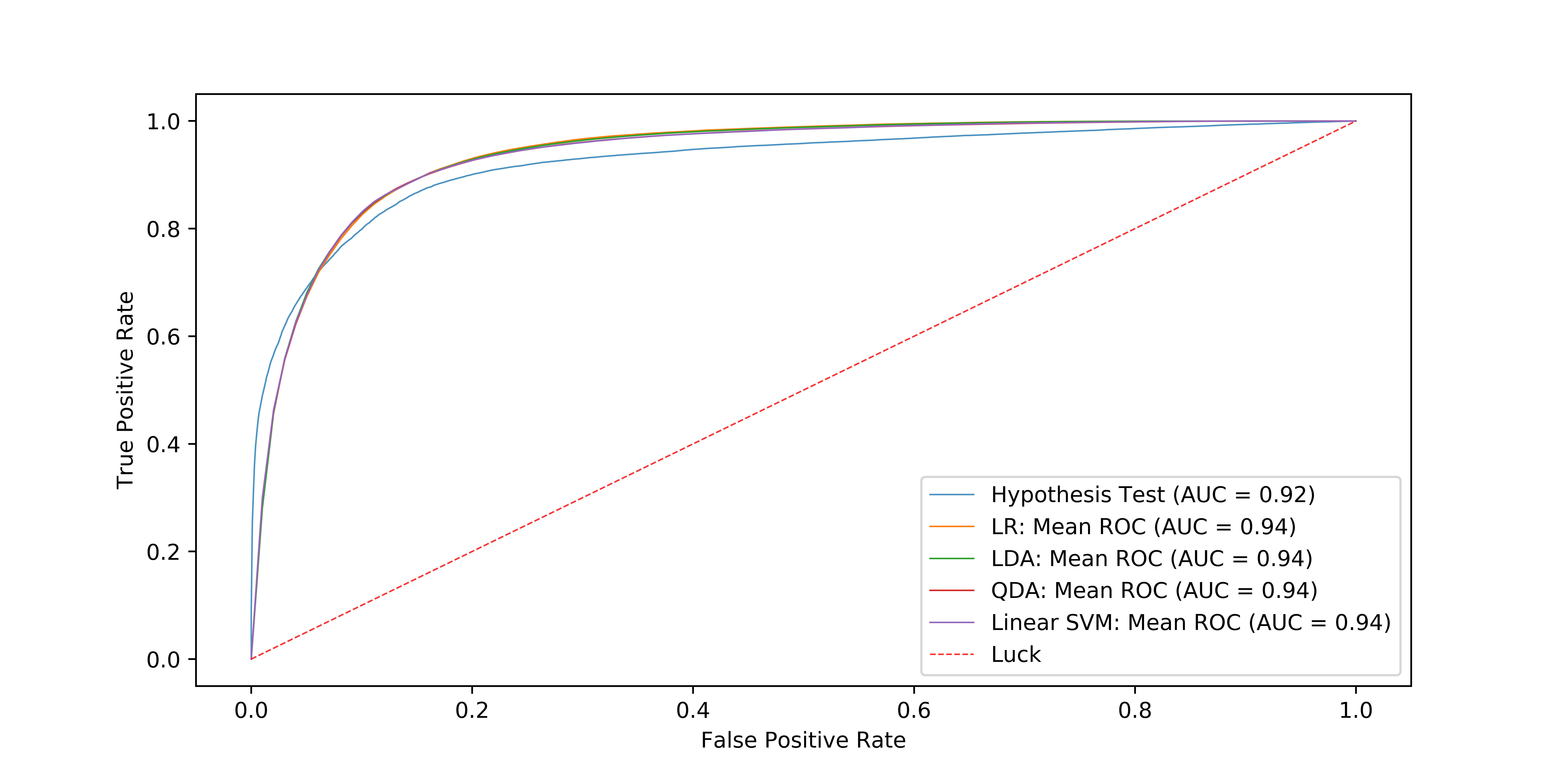}
}
\subfigure[medium SNR ($A_t=0.5$) ]{
\includegraphics[width=.5\textwidth]{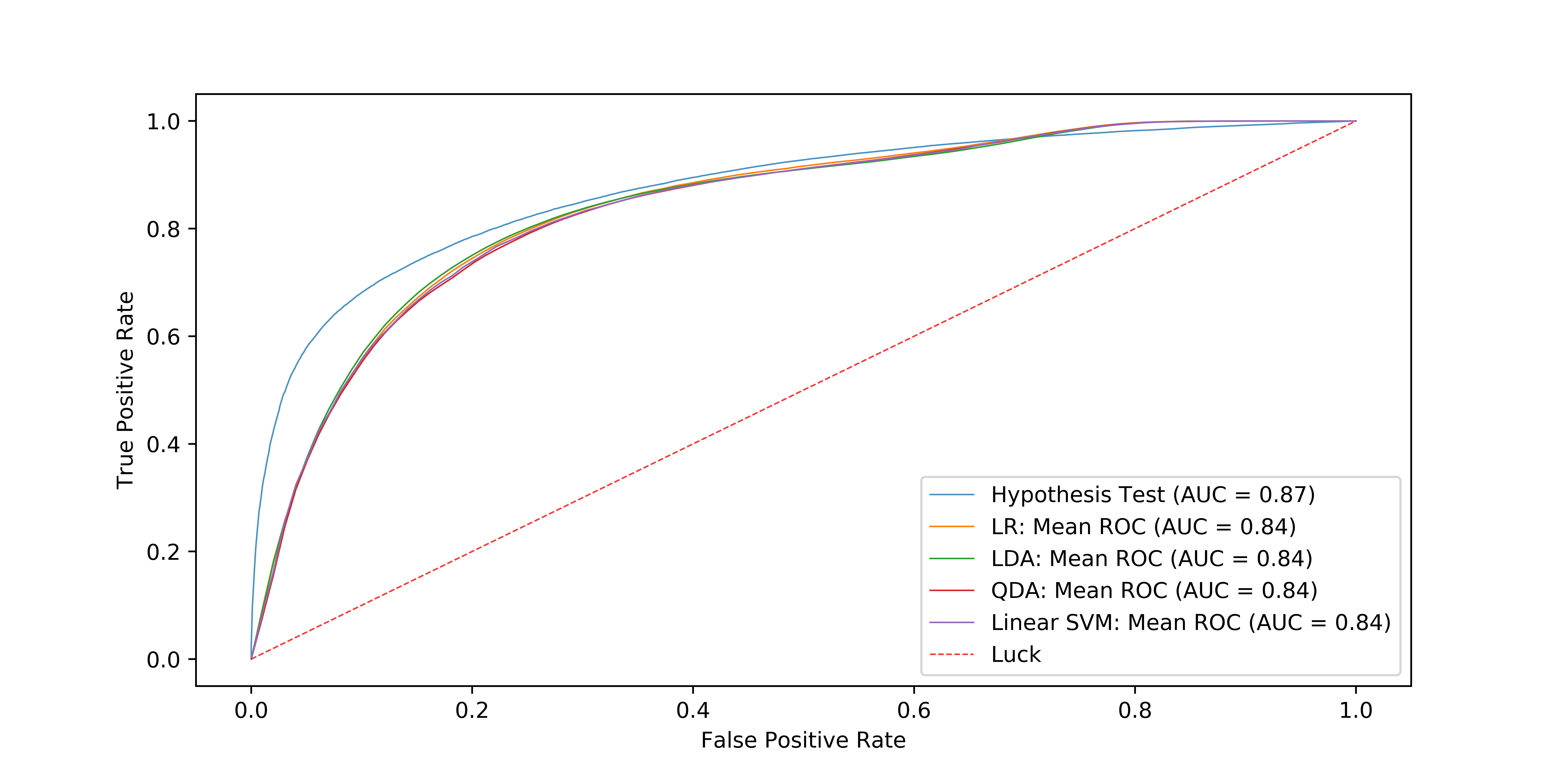}
}

\subfigure[high SNR ($A_t=0.6$) ]{
\includegraphics[width=.5\textwidth]{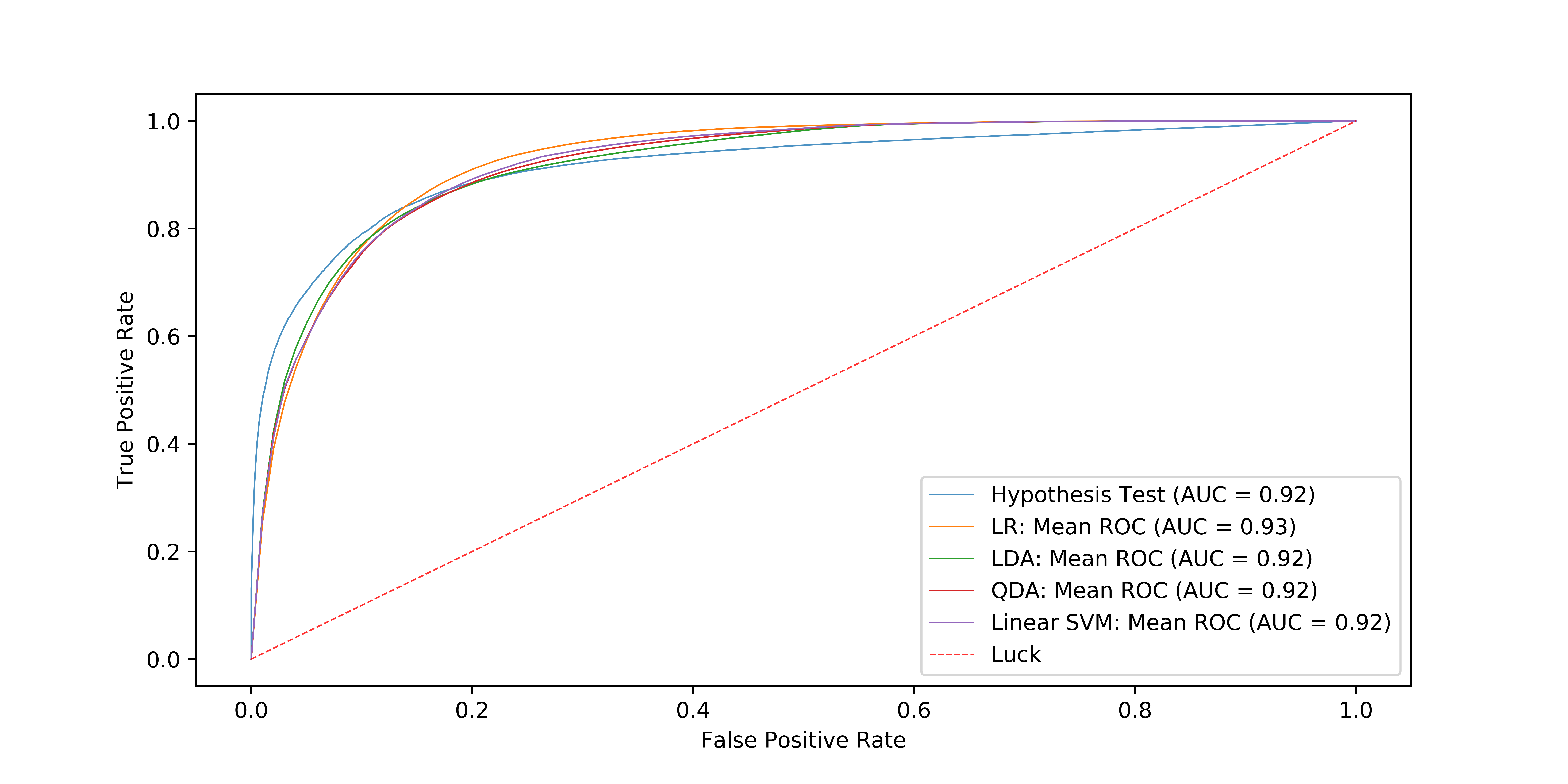}
}

\caption{Receiver operating characteristic (ROC) curves. AUC stands for area under the curve.}
\label{fig:ROC}
\end{figure}

\begin{table*}[]
    \centering
 
\caption{Performance of the BHT and the ML classifiers}   
\begin{tabular}{ |c|c|c|c|c|c|c|c|c|c|c|c|c|c|c|c| }
 \hline
 \textbf{Metric} & \multicolumn{5}{|c|}{\textbf{Low SNR ($A_t = 0.4$)}} & \multicolumn{5}{|c|}{\textbf{Medium SNR ($A_t = 0.5$)}} & \multicolumn{5}{|c|}{\textbf{High SNR ($A_t = 0.6$)}}\\
 \hline
   & LR    &SVM& BHT  & LDA & QDA & LR    &SVM& BHT  & LDA & QDA  & LR    &SVM& BHT  & LDA & QDA   \\ 
 \hline
  PFA  &  0.001   &0.092&0.132   &0.18 &0.19 &0.23 &0.2 &0.21 &0.38 &0.62 &0.163 &0.09 &0.15 &0.35 &0.42 \\ 
  \hline
 PMD  & 0.98   &0.1& 0.08  & 0.11&0.13 &0.41 &0.15 &0.17 &0.235 &0.05 &0.358 &0.135 &0.09 &0.23 &0.03 \\ 
 \hline
Accuracy (\%)  & 57.31    &88.24& 87.46  &84.51 &84.11 &68.26 &83.91 &81.21 &69.56 &66.90 &77.91 &87.38 &86.65 &72.82 &78.29 \\ 
 \hline
\end{tabular}
 
    \label{tab:P}
\end{table*}

%\begin{figure*}
%    \centering
%    \subfigure[]{\includegraphics[width=0.4\textwidth]{PFA_HSG_AMP0.4.eps}} 
%    \subfigure[]{\includegraphics[width=0.4\textwidth]{PFA_HSG_AMP0.5.eps}} 
%    \subfigure[]{\includegraphics[width=0.4\textwidth]{PFA_HSG_AMP0.6.eps}}
%    \caption{False Alarm }
%    \label{fig:PFA}
%\end{figure*}

%\begin{figure*}
%    \centering
%    \subfigure[]{\includegraphics[width=0.24\textwidth]{PMD_HSG_AMP0.4.eps}} 
%    \subfigure[]{\includegraphics[width=0.24\textwidth]{PMD_HSG_AMP0.5.eps}} 
%    \subfigure[]{\includegraphics[width=0.24\textwidth]{PMD_HSG_AMP0.6.eps}}
%    \caption{Missed Detection }
%    \label{fig:PMD}
%\end{figure*}

%\begin{figure*}
%    \centering
%    \subfigure[]{\includegraphics[width=0.24\textwidth]{ACC_HSG_AMP0.4.eps}} 
%    \subfigure[]{\includegraphics[width=0.24\textwidth]{ACC_HSG_AMP0.5.eps}} 
%    \subfigure[]{\includegraphics[width=0.24\textwidth]{ACC_HSG_AMP0.6.eps}}
%    \caption{Accuracy}
%    \label{fig:Acc}
%\end{figure*}

Receiver operating characteristic (ROC) curves are one popular metric to evaluate the performance of ML and statistical classifiers. An ROC curve plots the correct decision rate against the error rate, i.e., true positive rate (i.e., deciding NLOS correctly) vs. false alarm/positive rate (i.e., deciding NLOS while it was LOS). Fig. \ref{fig:ROC} shows the ROC curves for all the four ML classifiers as well as the BHT (an statistical classifier), for three different link conditions. We make the following observations. 1) At low false alarm rates, the BHT performs the best among all the classifiers. But then, there is a switching mechanism in force where beyond a certain false positive rate, the ML classifiers outperform the BHT. 2) To our surprise, an increase in SNR doesn't lead to a monotonous increase in the accuracy of all the proposed NLOS identification methods. This is probably due to residual effects of multipath, small-scale fading and additive noise, and calls for more measurements for each receiver position, and longer time-slot intervals so that we get more stable pathloss measurements due to increased averaging. 

%\textcolor{red}{Please check :Fig. \ref{} plots the ROC curve for the case when multiple features are collectively used to train the ML algorithms. For other features (other than pathloss), we generated synthetic data which is inline with the work of \cite{Yu:TVT:2009}. For the multiple features, it was hard to derive closed-form expressions for the error probabilities, so we have omitted it.}
 
Table II evaluates the NLOS identification performance of the BHT and the four ML classifiers based on the following three performance metrics: (a) probability of false alarm (PFA), (b) probability of missed detection (PMD), and (c) accuracy, where accuracy$=(1-$(PMD+PFA))$\times100$. We make the following observations. 1) The performance of the best-performing ML algorithm (i.e., the RBF-SVM classifier) is only slightly superior to the Neyman-Pearson-based BHT. That is, the RBF-SVM classifier (the best performing ML classifier) and the BHT achieve a maximum accuracy of 88.24\% and 87.46\% for low SNR, 83.91\% and 81.21\% for medium SNR, and 87.38\% and 86.65\% for high SNR. 2) Some ML classifiers (e.g., QDA) perform worst for one error type (i.e., false alarm rate) but perform best for the other error type (i.e. missed detection rate), and vice versa (e.g., LR). However, BHT and RBF-SVM are efficient in the sense that they minimize both error types simultaneously. 3) Again, to our surprise, an increase in SNR doesn't lead to a monotonous increase in the accuracy of all the proposed NLOS identification methods (due to insufficient averaging while obtaining pathloss measurements). 

%\vfill\pagebreak

%\section{NLOS Prediction via Multiple Physical-Layer Features}
%optimal weight selection via convex optimization, densities of measured features are not always normal, multi-dimensional classification which is where ML techniques become handy because of lack of analytical tractability. 

%Let $\mathbf{z}=[z_{PL},z_{\phi_A},z_{\phi_E},z_{TOA}]^T$. Assuming all the measurements are independent and Gaussian, we have the following test:
%\begin{equation}
%	\label{eq:H0H1_1_2}
%	 T=\mathbf{z}^T\Sigma^{-1}\mathbf{z} \underset{H_0}{\overset{H_1}{\gtrless}} \zeta 
%\end{equation}
%where $\Sigma=\text{diag}[\sigma_{PL}^2,\sigma_{\phi_A}^2,\sigma_{\phi_E}^2,\sigma_{TOA}^2]^T$.One can see that the test statistic $T$ follows central (non-central) chi-squared distribution under $T|H_0$ ($T|H_1$), i.e., $T|H_0\sim \chi^2(4)$, and $T|H_1\sim \chi^2(4)$. Then, the two error probabilities are given as: $P_{fa}=Pr(\mathbf{z}>\zeta|H_0)$, and $P_{d}=1-P_{md}=1-Pr(\mathbf{z}<\zeta|H_1)$.

\section{Conclusion}

This paper conducted an experimental study on the problem of LOS/NLOS classification in an indoor environment. We used a pair of NI 2901 USRP SDRs in a large hall (with receive SDR moving on a 1D grid) in order to construct a dataset of pathloss measurements (for both LOS and NLOS scenarios). We utilized our custom dataset to estimate the pathloss parameters (i.e., pathloss exponent) using the least-squares method, and later, utilized the parameterized pathloss model to construct a binary hypothesis test for NLOS identification. Further, noting that the pathloss measurements slightly deviate from the Gaussian distribution, we passed our custom dataset to four ML algorithms, i.e., linear and radial basis function SVM, LDA, QDA, and LR. We observed that the best-performing ML algorithm (i.e., RBF-SVM) marginally outperformed the Neyman-Pearson-based binary hypothesis test. 

As for the future work, we note that the ML-based techniques are environment-specific, i.e., if the environment changes, we need to train the ML algorithms again. So, one promising future direction is to design reinforcement/online learning methods for NLOS identification.

\bibliographystyle{IEEEtran}
\bibliography{refs}

\end{document}